\documentclass[aps,prl,floatfix,twocolumn,preprintnumbers,amsmath,amssymb,groupedaddress,showpacs,showkeys]{revtex4-1}

\usepackage{graphicx} 
\usepackage{dcolumn} 
\usepackage{bm} 
\usepackage{color} 
\usepackage{amsmath,amssymb}
\usepackage{psfrag}
\usepackage{mathrsfs}

\newcommand{\ua}{\uparrow}
\newcommand{\da}{\downarrow}

\begin{document}

\title{Resonant scattering by magnetic impurities as a model for spin relaxation in bilayer graphene}
\author{Denis Kochan, Susanne Irmer, Martin Gmitra and Jaroslav Fabian}
\affiliation{Institute for Theoretical Physics, University of Regensburg, 93040 Regensburg, Germany\\
}
\begin{abstract}
We propose that the observed spin-relaxation in bilayer graphene is due to resonant scattering by
magnetic impurities. We analyze a resonant scattering model due to adatoms on both dimer and non-dimer sites,
finding that only the former give narrow resonances at the charge neutrality point.
Opposite to single-layer graphene, the measured spin-relaxation rate in graphene bilayer increases with carrier density.
Although it has been commonly argued that a different mechanism must be at play for the two structures, our model explains
this behavior rather naturally in terms of different broadening scales for the same underlying resonant processes.
Not only our results---using robust and first-principles inspired parameters---agree with experiment, they
also predict an experimentally testable sharp decrease of the spin-relaxation rate at high
carrier densities.
\end{abstract}

\date{\today}
\maketitle

Understanding spin-relaxation is essential for designing spintronics devices~\cite{Zutic2004:RMP,Fabian2007:APS}.~Unfortunately, spin-relaxation in graphene structures has been a baffling problem~\cite{Han2014:NatCom}. While experiments in both single layer
graphene (SLG)~\cite{Tombros2007:N, Tombros2008:PRL, Pi2010:PRL, Yang2011:PRL, Han2011:PRL, Avsar2011:NanoLett, Jo2011:PRB, Mani2012:NC}
and bilayer graphene (BLG)~\cite{Yang2011:PRL, Han2011:PRL} yield spin lifetimes
on the $100 - 1000$\,ps time scale (the highest values achieved in
graphene/h-BN structures~\cite{Guimaraes2014:PRL, Drogeler2014:NL}),
theories based on realistic spin-orbit coupling and transport parameters predict lifetimes
on the order of microseconds~\cite{Huertas2006, Dora2010:EPL, Jeong2011:PRB, Dugaev2011:PRB,%
Ertler2009:PRB, Zhang2011:PRB, Ochoa2012:PRL, Diez2012:PRB, Wang2013:PRB, Fedorov2013:PRL, Tuan2014:NP}.

While the magnitudes of the spin-relaxation rates of SLG and BLG are similar, the dependence
of the rates on the electron density is opposite in the two systems. In SLG the spin-relaxation
rate decreases with increasing the carrier density~\cite{Tombros2008:PRL, Pi2010:PRL, Yang2011:PRL, Han2011:PRL},
in BLG the spin-relaxation rate increases~\cite{Yang2011:PRL, Han2011:PRL}. Since
the diffusivity in the investigated samples decreases with increasing the electron density,
it has been a common practice to assign two different mechanisms to both structures:
the Elliott-Yafet mechanism~\cite{Elliott1954:PR, Yafet1963} to SLG~\cite{Tombros2008:PRL, Pi2010:PRL, Avsar2011:NanoLett, Jo2011:PRB} and
Dyakonov-Perel mechanism~\cite{Dyakonov1972:SovPhys} to BLG~\cite{Yang2011:PRL, Han2011:PRL, Avsar2011:NanoLett}.

The main problem with that assignment is quantitative. Spin-orbit coupling in graphene~\cite{Gmitra2009:PRB}
is too weak to yield such a small spin-relaxation time. An explicit first-principles calculation~\cite{Fedorov2013:PRL}
predicts that one would need 0.1~\% of adatoms to give $100$\,ps
spin lifetime. Recently a new mechanism for SLG was proposed~\cite{Kochan2014:PRL}
(see also Ref.~\cite{Soriano2015:A}), based on resonant scattering off local magnetic moments. It gives the observed spin-relaxation times with as little as 1\,ppm of local magnetic moments and also agrees with the
experimental behavior for SLG of decreasing the spin-relaxation rate with increasing electron
density. Where do these local moments come from? It was theoretically predicted that adatoms
such as hydrogen~\cite{Duplock2004:PRL, Yazyev2010:RPP}, but also chemisorbed organic molecules
\cite{Santos2012:NJP} can be responsible. Experimentally it was demonstrated
that hydrogen adatoms indeed induce local moments~\cite{McCreary2012:PRL,Birkner2013:PRB}, but even untreated graphene flakes
were shown to exhibit 20 ppm spin 1/2 paramagnetic moments~\cite{Nair2012:NP}. The most natural candidates for resonant magnetic
scatterers appear to be polymer residues from different fabrication steps of graphene devices.
Our mechanism is also in line with mesoscopic transport experiments~\cite{Kozikov2012:PRB, Lundeberg2013:PRL} which found a strong evidence for the local magnetic moments in the dephasing rates. Our
theory does not work for high adatom concentrations (say, above 0.1\%), at which the induced
magnetic moments seem to form a fluctuating magnetic-field network~\cite{McCreary2012:PRL}.

\begin{figure*}
\begin{center}
\includegraphics[width=1.9\columnwidth,angle=0]{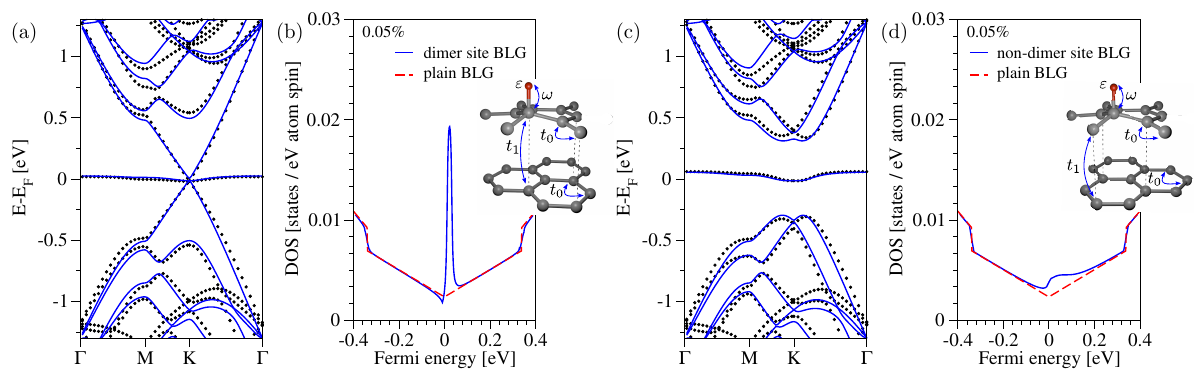}
\end{center}
\vspace*{0.2cm}
\caption{(Color online) Calculated electronic structure of bilayer graphene with hydrogen adatoms. Panels~(a)~and~(b) are for dimer adatoms, (c)~and~(d) for non-dimer ones.
In (a)~and~(c) we plot the electronic band structures: dotted lines are spin-unpolarized first-principles calculations using a 7$\times$7 supercell, while solid lines are tight-binding fits as described in text. Panels~(b)~and~(d) show unperturbed, $\varrho_{0}^{+}(E)+\varrho_{0}^{-}(E)$, and perturbed, $\mathcal{R}_{\mathrm{C}}(E)$ (with adatom concentration of $\eta=0.05\%$), DOS per atom and spin.
Dimer adatoms (b) show a narrow resonant peak near the charge neutrality point at $E_\mathrm{res}\simeq 22.5$\,meV with the full width at half maximum $\Gamma\simeq 8.4\,$meV. Non-dimer adatoms (d) induce a broad resonance at $E_\mathrm{res}\simeq 26.1$\,meV with $\Gamma\simeq 165.2$\,meV. For plotting DOS we perform running averages of 20~meV.
Insets: schemes of the tight-binding model Hamiltonian, $H_{0}+H'$, Eqs.~(\ref{eq:BLG-hamiltonian})~and~(\ref{eq:perturbation-hamiltonian}).}
\label{Fig:1}
\end{figure*}

In this letter we propose that the
spin-relaxation in BLG is caused by the same mechanism of resonant magnetic scatterers. We show that
(i) adatoms (we model specifically hydrogen) on dimer and non-dimer sites of BLG give different resonance energies and resonance widths,
(ii) the calculated spin-relaxation times are in quantitative agreement with experiment,
(iii) the opposite trends of the spin-relaxation rate in SLG and BLG are due to different scales of the
energy fluctuations (caused by electron-hole puddles) in the two structures, reflecting their different density of states (DOS),
(iv) the spin-relaxation rate in BLG should reverse its trend and decrease with increasing electron density
at high densities, making an experimentally verifiable prediction.
As in SLG, also in BLG resonant magnetic scatterers
are spin hot spots~\cite{Fabian1998:PRL}: affecting spin but not momentum relaxation.

\paragraph{Model Hamiltonian.}
We consider a single adatom on AB stacked bilayer graphene sitting on either a dimer or a non-dimer position. The full model Hamiltonian is
$H_0 + H'$, where
\begin{align}
H_{0}&=-t_0\sum_{\langle m,n\rangle \sigma \atop \lambda\in\{\mathrm{t},\mathrm{b}\}}|a_{m\sigma}^{\lambda}\rangle\,\langle b_{n\sigma}^{\lambda}|
+t_1\sum\limits_{m\sigma} |a_{m\sigma}^{\mathrm{t}}\rangle\,\langle b_{m\sigma}^{\mathrm{b}}| + {\rm h.c.} \label{eq:BLG-hamiltonian},
\end{align}
is the unperturbed BLG Hamiltonian with intralayer nearest-neighbor hopping $t_0=2.6\,$eV,
and direct interlayer hopping $t_1=0.34\,$eV~\cite{Konschuh2012:PRB}. We neglect indirect interlayer parameters which
yield fine features of the energy bands (such as warping and electron-hole asymmetry) as unimportant for our purposes.
The first sum runs over $\langle m,n\rangle$ nearest neighbors in the top ($\lambda=\mathrm{t}$) and bottom ($\lambda=\mathrm{b}$) layers.
The second sum runs over the $a$-sites of the A sublattice of the top layer and $b$-sites of the B sublattice of the
bottom layer. State $|c_{m\sigma}^{\lambda}\rangle$ represents the spin $\sigma$ carbon $2p_z$ orbital on sublattice $c=\{a,b\}$ and site $m$ in
layer $\lambda$. The eigenstates of $H_0$ are
\cite{Johnson1973:PRB, McCann2013:RepProgPhys}
\begin{equation}\label{eq:BLG-eigenenergy}
\varepsilon_\alpha^\mu(\mathbf{k})=\frac{\alpha}{2}\Bigl(\mu t_1+\sqrt{t_1^2+4t_0^2|f(\mathbf{k})|^2}\,\Bigr)\,,
\end{equation}
where the index $\alpha$ labels the conduction ($\alpha=+$) and valence ($\alpha=-$) bands, and
$\mu$ stands for the high ($\mu=+$) and low ($\mu=-$) energy bands with respect to the charge neutrality point, for $f(\mathbf{k})$ see \cite{SM}.

\begin{figure*}
\begin{center}
\includegraphics[width=1.85\columnwidth,angle=0]{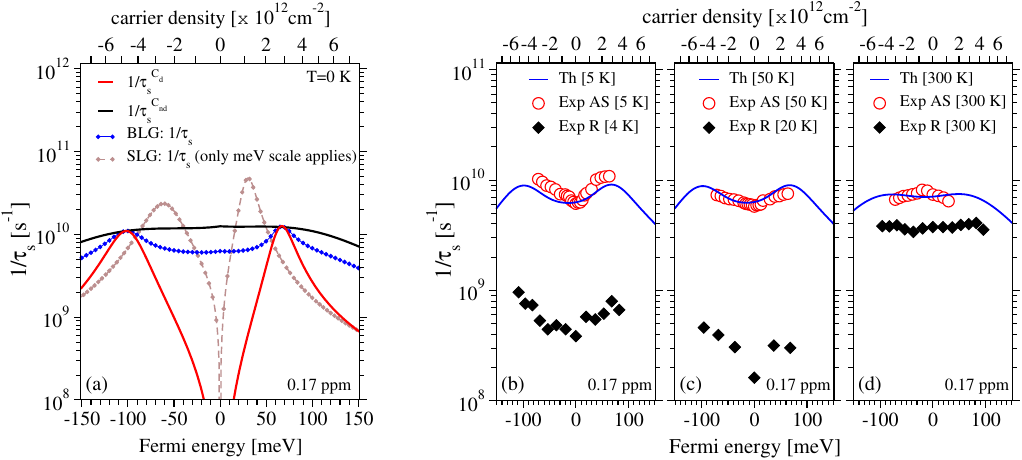}\\
\end{center}
\vspace*{0.2cm}
\caption{(Color online) Calculated spin-relaxation rates $1/\tau_s$ as a function of energy (carrier-density) for impurity concentration $\eta=0.17\,$ppm.
(a) Zero temperature, unbroadened, rates for dimer (red solid) and non-dimer (black solid) adatoms, as well as the resulting average $1/\tau_s$ (blue dotted).
For reference the SLG calculation is also shown (gray dashed-dotted). (b)--(d) Spin-relaxation rates for three representative temperatures. Theoretical data
(blue solid) are broadened, simulating the presence of electron-hole puddles, with a Fermi level smearing of $23\,$meV. Circles and diamonds represent data points from Aachen-Singapore~(AS)~\cite{Yang2011:PRL} and Riverside~(R)~\cite{Han2011:PRL} experiments, respectively. The two shoulders (spin-relaxation
edges) at $\pm 100$~meV are exchange-split resonances. At high carrier densities the model predicts a decrease of the spin-relaxation rate.}
\label{Fig:2}
\end{figure*}

We place the adatom on the top layer either on a dimer site $\mathrm{C}_{\mathrm{d}}$, which is an $a^{\mathrm{t}}$ site, or
on a nondimer site $\mathrm{C}_{\mathrm{nd}}$, which is a $b^{\mathrm{t}}$ site. The adatom also carries a local magnetic
moment coupled with the electron spins via the exchange interaction $J$. The Hamiltonian $H'$ is~\cite{Kochan2014:PRL},
\begin{equation}\label{eq:perturbation-hamiltonian}
H'=\sum\limits_{\sigma}\varepsilon|h_{\sigma}\rangle\,\langle h_{\sigma}|+
\omega\bigl(|h_{\sigma}\rangle\,\langle c_{\sigma}|+|c_{\sigma}\rangle\,\langle h_{\sigma}|\bigr)
-J\,\hat{\mathbf{s}}\cdot\hat{\mathbf{S}},
\end{equation}
where $|h_{\sigma}\rangle$ is the adatom orbital with spin $\sigma$. This orbital has on-site energy $\varepsilon$ and
is connected to the site $\mathrm{C}$ on the bilayer with hopping energy $\omega$. The spin operators $\hat{\mathbf{s}}$
and $\hat{\mathbf{S}}$, which are the Pauli matrices in the corresponding spinor spaces, are for itinerant electrons and local
magnetic moments, respectively.

To obtain realistic parameters for the adatom Hamiltonian $H'$, we performed first-principles calculations with {\sc{Quantum ESPRESSO}} \cite{Giannozzi2009:JPCM} using a $7\times 7$ graphene supercell with a single hydrogen
adatom. In agreement with previous studies~\cite{Moaied2014:PRB} we found that hydrogen on both dimer and non-dimer sites induces local magnetic moments of 1~Bohr magneton per unit cell.
However, for fitting the orbital parameters of $H'$, namely $\varepsilon$ and $\omega$,
we used the spin-unpolarized first-principles band structure and set $J=0$ in the tight-binding calculation.
For the dimer site we select $\varepsilon=0.25\,$eV and $\omega=6.5\,$eV, while for the non-dimer one
$\varepsilon=0.35\,$eV and $\omega=5.5\,$eV. Figures~\ref{Fig:1}(a)~and~(c) show that the fits are very good.
However, these fitted parameters are not unique, as a larger neighborhood of values offers a comfortable agreement
with first-principles data. We have selected the values which are close to the uniquely fitted SLG orbital parameters. For the exchange coupling we take the same (unbiased) value as for SLG~\cite{Kochan2014:PRL}, $J = -0.4$\,eV.

\paragraph{Resonant scattering.}
We transform the adatom Hamiltonian $H'$ into the singlet ($\ell=0$) and triplet ($\ell=1$) basis $|c_{\ell,m_\ell}\rangle=|c\rangle\otimes|{\ell,m_\ell}\rangle$ [label $m_\ell=-\ell,\dots,\ell$ is the total spin projection] and eliminate the adatom orbital $|h\rangle$ by downfolding.
This gives the energy dependent perturbation
$H'(E)=\sum_{\ell,m_\ell} {V}_\ell(E)\ |c_{\ell,m_\ell}\rangle\,\langle c_{\ell,m_\ell}|$, allowing us to analytically calculate
the T-matrix, $\mathrm{T}(E)=\sum_{\ell,m_\ell} \mathrm{T}_\ell(E)\ |c_{\ell,m_\ell}\rangle\,\langle c_{\ell,m_\ell}|$, where
\begin{align}
{V}_\ell(E)=\tfrac{\omega^2}{E-\varepsilon+(4\ell-3)J}\,,\ \ \
\mathrm{T}_\ell(E)=\tfrac{{V}_\ell(E)}{1-{V}_\ell(E)\,G_{\mathrm{C}}(E)}\,.
\end{align}
Here $G_{\mathrm{C}}(E)\equiv\Lambda_{\mathrm{C}}(E)-i\pi\nu_{\mathrm{C}}(E)$ is the $\mathrm{C}$-site projected Green's function per atom and spin of the
unperturbed BLG with
\begin{align}
\Lambda_{\mathrm{C}}(E)&=\tfrac{E}{2D^2}\ln\Bigl|\tfrac{E^2(E^2-t_1^2)}{(D^2-E^2)^2}\Bigr|+\tfrac{t_1\Delta_\mathrm{C}}{2D^2}\ln\Bigl|\tfrac{E+t_1}{E-t_1}\Bigr|\,,\\
\nu_{\mathrm{C}}(E)&=\sum\limits_{{\mu=\pm}}\tfrac{|E|-\mu\Delta_\mathrm{C}t_1}{2D^2}\,\Theta\bigl(D-|E|\bigr)\Theta\bigl(|E|-\mu t_1\bigr)\,,\label{eq:projected DOS}
\end{align}
where $D=\sqrt{\sqrt{3}\pi}t_0\simeq 6\,$eV is the effective BLG bandwidth and
$\Delta_\mathrm{C}$ equals zero for $\mathrm{C_d}$-site and one for $\mathrm{C_{nd}}$-site, respectively.

We first analyze orbital resonances of $H'$ (set $J=0$) by plotting in Fig.~\ref{Fig:1}(b)~and~(d) the perturbed DOS per atom and spin,
$\mathcal{R}_{\mathrm{C}}(E)=\sum_{\mu=\pm}\varrho_0^\mu(E)-{(\eta/\pi)}\,\mathrm{Im}\,\bigl\{\bigl[-\tfrac{d}{dE}G_{\mathrm{C}}(E)\bigr]\,\mathrm{T}_\ell(E,J=0)\bigr\}$, where $\eta$ is the adatom concentration per carbon atom and
$\varrho_0^\mu(E)=(2|E|-\mu t_1)/(4D^2)\,\Theta\bigl(D-|E|\bigr)\,\Theta\bigl(|E|-\mu t_1\bigr)$
is the unperturbed bilayer DOS per atom and spin for the high ($\mu=+$) and low ($\mu=-$) energy band, for details see~\cite{SM}.
As seen from Fig.~\ref{Fig:1}(b), the dimer site hydrogen chemisorption induces a pronounced narrow resonance near the
charge neutrality point. In contrast, non-dimer adatoms, see Fig.~\ref{Fig:1}(d), give a broad and shallow resonance.
This striking difference is explained by considering the character of the resonance states.
 In a monolayer graphene an adatom on the A site induces a resonance state which is localized mainly on B sublattice. Thus, an adatom on a dimer site induces a resonance state which is spread mainly on the non-dimer sublattice and hybridizes only little with the other layer, keeping the resonance narrow. If the adatom is on a non-dimer site, the resonance state is mainly on the dimer sublattice which couples to the other layer, causing a leakage of the state and broadening of the resonance. The same behavior
is seen in vacancy models \cite{Castro2010:PRL}.

\paragraph{Spin-flip scattering and spin-relaxation rate.}
The T-matrix allows us to compute the spin-flip rate for a single scattering event by adatom at site $\mathrm{C}$ (dimer
or non-dimer) $|\mathbf{k}^{\mu}(E),\ua\rangle\rightarrow|\mathbf{q}^{\nu}(E'),\da\rangle$ between bands
$\mu$ and $\nu$~\cite{note1},
\begin{align}\label{eq:single-rate}
\hspace{-2mm}W^{\mathrm{C}}_{\mathbf{k}^{\mu}\ua,\mathbf{q}^{\nu}\da}=
\frac{2\pi}{\hbar}\,\eta^2\,
f_{\ua,\da}^{\mathrm{C}}(E)\,
\mathrm{P}_{\mathrm{C}}^{\mu}(E)\mathrm{P}_{\mathrm{C}}^{\nu}(E')\,\delta(E-E')\,,
\end{align}
where the site and band dependent projections
$\mathrm{P}_\mathrm{C}^\mu(E)=2(|E|-\mu\Delta_\mathrm{C}t_1)/(2|E|-\mu t_1)\,\Theta\bigl(D-|E|\bigr)\,\Theta(|E|-\mu t_1)$, see also~\cite{SM}. The exchange-induced spin-flip function is
\begin{equation}\label{eq:spin-flip-function}
f_{\ua,\da}^{\mathrm{C}}(E)=\Biggl|\,\sum\limits_{\ell=0,1}\frac{\bigl(\ell-\tfrac{1}{2}\bigr)\,{V}_\ell(E)}{1-{V}_\ell(E)\,G_{\mathrm{C}}(E)}\,\Biggr|^2\,.
\end{equation}
The spin-flip rate does not depend on the relative orientation of $\mathbf{k}$ and $\mathbf{q}$, since the energy dispersion in our model has rotational symmetry.
However, the spin-flip rate is very different for dimer and non-dimer adatoms.

To obtain the spin-relaxation rate $1/\tau_s^{\mathrm{C}}$ we sum over different partial rates and obtain
\begin{align}\label{eq:Relax-Rate}
\hspace{-1mm}\frac{1}{\tau_s^{\mathrm{C}}}=\eta\,\frac{2\pi}{\hbar}f_{\ua,\da}^{\mathrm{C}}(E)
\frac{\left[\mathrm{P}_{\mathrm{C}}^{+}(E)\varrho_0^{+}(E)+\mathrm{P}_{\mathrm{C}}^{-}(E)\varrho_0^{-}(E)\right]^2}{\varrho_0^{+}(E)+\varrho_0^{-}(E)}
\,,
\end{align}
where the labels $+$ and $-$ denote BLG high and low energy bands entering the definitions of $\mathrm{P}_{\mathrm{C}}^\mu$ and $\varrho_0^\mu$ given in the text.
To get the final spin-relaxation rate we take an unbiased average over the dimer and non-dimer sites,
$1/\tau_s\equiv 1/\bigl(2\tau_s^{\mathrm{C_d}}\bigr)+1/\bigl(2\tau_s^{\mathrm{C_{nd}}}\bigr)$. This is plotted in Fig.~\ref{Fig:2}(a) and compared with SLG.
Two pronounced shoulders---we call them \emph{spin-relaxation edges}---in $1/\tau_s^{\mathrm{C_d}}$ emerge from the exchange splitting of the orbital resonance seen in DOS at Fig.~\ref{Fig:1}(b), just like
for SLG, although the peaks in BLG are more separated due to the energy renormalization by the interlayer coupling.
In contrast, non-dimer adatoms show a rather flat behavior with respect to the energy, reflecting the broad
resonance of the perturbed DOS, Fig.~\ref{Fig:1}(d). Non-dimer adatoms still induce a large
$1/\tau_s$ since they strongly perturb the low-energy states which are localized on the non-dimer sites.
This behavior is encoded in the low-energy site projection $P_{\mathrm{C}}^{-}(E)$, in Eq.~(\ref{eq:Relax-Rate}),
which is at low energies much larger for non-dimer than for dimer adatoms.

\paragraph{Comparison with experiments and contrasting single and bilayer graphene.} Comparison with experiments requires temperature and electron-hole
puddles broadening of $1/\tau_s$. Temperature broadening is due to population smearing, $(-{\partial f_0}/{\partial E})$,
where $f_0$ is the Fermi-Dirac distribution. The puddle broadening is modeled as a convolution with
a Gaussian kernel of standard deviation $\sigma_{\mathrm{br}}$. For bilayer we use $\sigma_{\mathrm{br}}\simeq 23\,$meV, which corresponds
to density fluctuations $\Delta n$ of $8.5\times 10^{11}/\mathrm{cm^2}$, following experimental estimates~\cite{Zou2010:PRB}.
In~Figs.~\ref{Fig:2}(b)-(d) we present the main results of this paper, the fully broadened spin-relaxation rates compared with
Aachen-Singapore~(AS)~\cite{Yang2011:PRL} and Riverside~(R)~\cite{Han2011:PRL} experiments. Clearly the two experiments are somewhat at odds,
but they display consistent behavior at low temperatures. We adjust the local moment concentration to describe the AS data,
obtaining $\eta=0.17\,$ppm. All other parameters are as obtained
from the orbital fits. The agreement at low temperatures is especially good. At high temperatures the overall shapes differ,
but the two experiments differ as well. This experimental discrepancy further underlines the extrinsic character of the spin-relaxation
in BLG. It is likely that the relative population of dimer and non-dimer adatoms changes with temperature,
differently in different samples, reflecting the idiosyncrasy of the experimental data.
However, our calculation gives a rather robust prediction at low temperatures:
{\it at high carrier densities, above the spin-relaxation edge at about $5\times 10^{12}/\mathrm{cm^2}$, the spin-relaxation rates should start to decrease.}

\begin{figure}
\begin{center}
\includegraphics[width=0.9\columnwidth,angle=0]{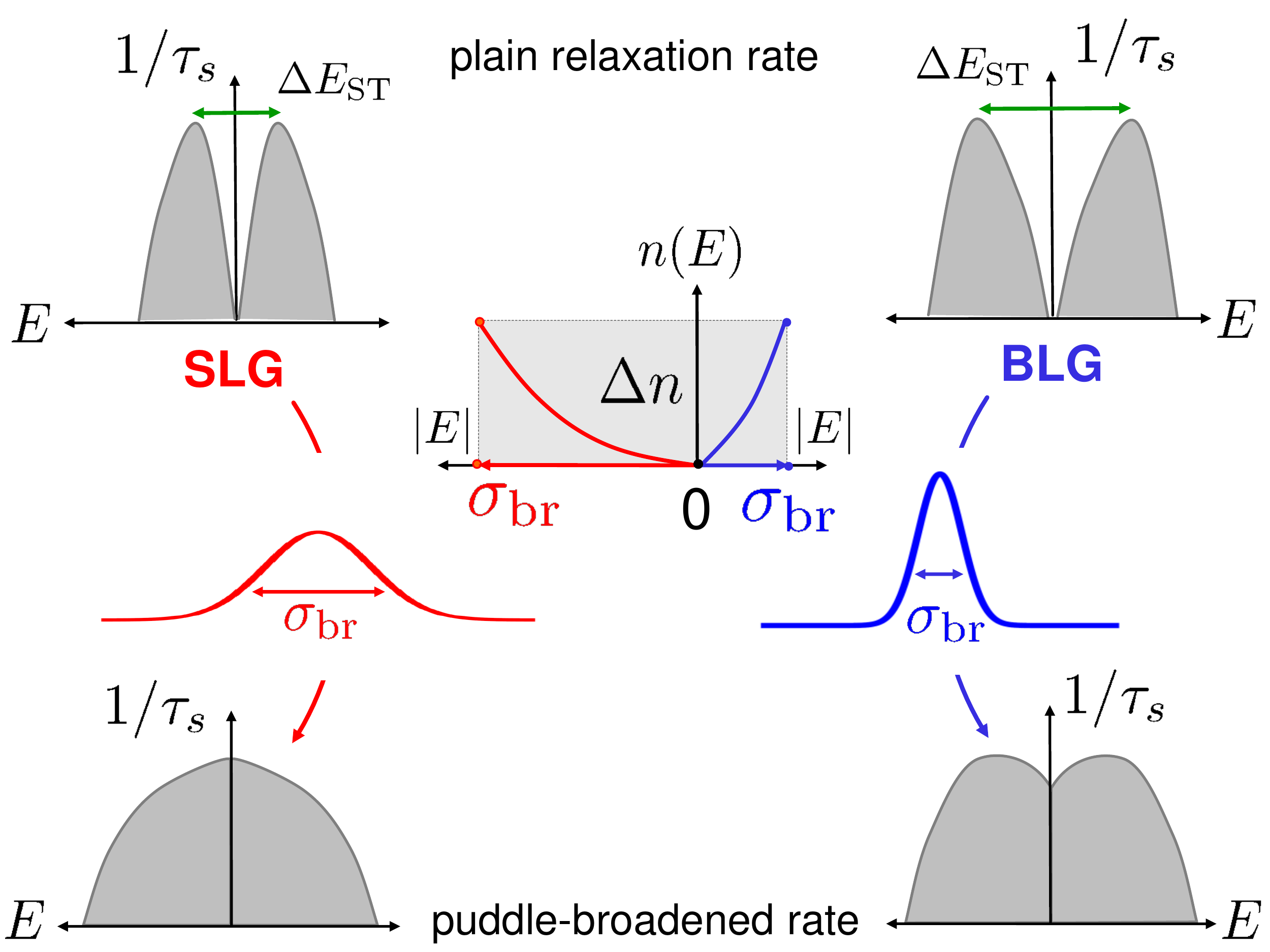}\\
\end{center}
\vspace*{0.2cm}
\caption{(Color online) The effect of electron-hole puddles on spin-relaxation in SLG and BLG. From top to bottom:
The spin-relaxation rate exhibits two resonance peaks due to singlet-triplet splitting $\Delta E_{\mathrm{ST}}$. The
splitting of the peaks is greater in BLG. The peaks are broadened
by temperature and carrier density fluctuations $\Delta n$ which is very different for SLG and BLG,
due to their different DOS. For a given temperature and density fluctuation $\Delta n$ the energy smearing in SLG
$\sigma_{\mathrm{br}}\simeq\Delta E_{\mathrm{ST}}$, while in BLG $\sigma_{\mathrm{br}}\ll\Delta E_{\mathrm{ST}}$.
After broadening the spin-relaxation rate around the charge neutrality point in SLG has the opposite trend as the unbroadened rate. In BLG
the original trend is preserved.}
\label{Fig:3}
\end{figure}

Perhaps the most pressing remaining question is: Given the same resonant spin-relaxation mechanism for single and bilayer graphene,
why do their spin-relaxation rates have the opposite trends as functions of charge density~\cite{Yang2011:PRL, Han2011:PRL}?
Our mechanism offers a natural, and perhaps mundane answer: electron-hole
puddles. At low temperatures and in the absence of density fluctuations the two structures should exhibit the same trend, namely,
an increase of the spin-relaxation rate going away from the charge neutrality point. In SLG the behavior is exactly opposite. The reason is offered in Fig.~\ref{Fig:3}.
In SLG the carrier density
fluctuations lead to a large Fermi energy smearing ($\sigma_{\rm br}=91\,$meV versus $\sigma_{\rm br}=23\,$meV in BLG for the same carrier density fluctuation
$\Delta n$ of $8.5\times 10^{11}/\mathrm{cm^2}$).
Averaging over the Fermi energy of the singlet-triplet split spin-relaxation peaks then inverts the
shape of the spin-relaxation rate around the Dirac point. In bilayer, due to its greater density of states, the energy broadening is much more
modest, and the experiments (unless their samples would exhibit large variations of the electronic densities) find the behavior as expected for
an unbroadened system.
Figure~\ref{Fig:3} also shows the origin of the spin-relaxation edge and the robustness of
our prediction of the decrease of the spin-relaxation rate at greater electron densities. At high temperatures (above
100~K), it is enough to invoke thermal broadening to see the trend reversal even in ultraclean SLG, as its resonance peaks
are closer than those in BLG \cite{note1}. The picture given in
Fig.~\ref{Fig:3} could be used to analyze experimental trends in spin-relaxation in both SLG and BLG.

In conclusion, we showed that resonant scattering by local magnetic moments quantitatively accounts for the experimental data. This spin-relaxation mechanism also explains the apparently striking opposite behavior
of the measured spin-relaxation of SLG and BLG, offering a real alternative to quantitatively unsubstantiated
but often made assignment of the two distinct trends as Elliott-Yafet and Dyakonov-Perel. Finally, our model makes a specific prediction of reversing the
increase of the spin-relaxation rate in graphene bilayer with increasing carrier density, at high densities, accessible experimentally.

We thank B.~Beschoten and R.~Kawakami for providing us with their experimental data.
This work was supported by DFG~SFB~689 and GRK~1570, and by the EU~Seventh~Framework~Programme under Grant~Agreement~No.~604391~Graphene~Flagship.
\nocite{Perdew1996:PRL,Grimme2006:JCC,Barone2009:JCC,Vanderbilt1990:PRB}

\bibliography{grp_adatom}

\pagestyle{empty}
\begin{widetext}
\includegraphics[width=1.0\columnwidth,angle=0]{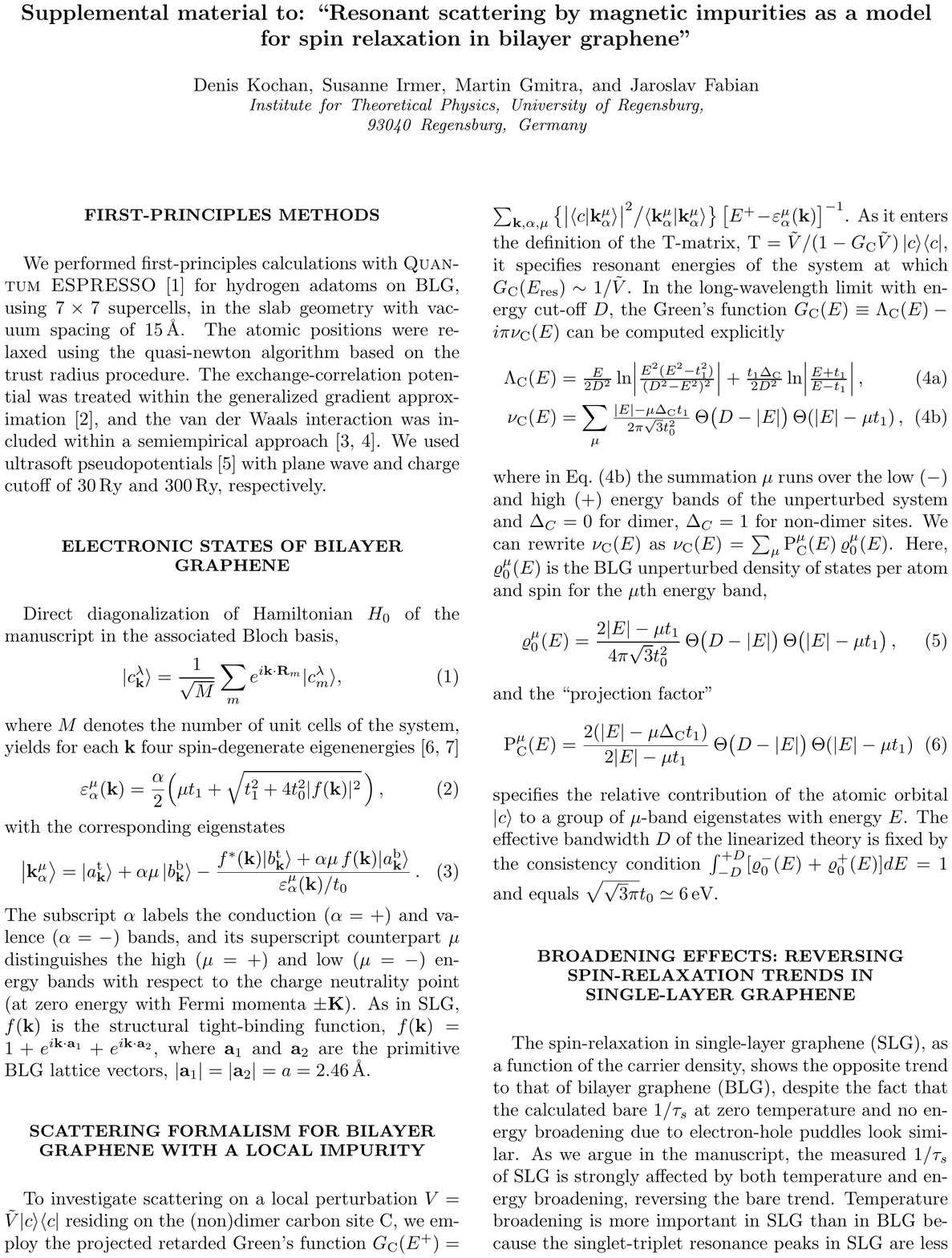}
\includegraphics[width=1.0\columnwidth,angle=0]{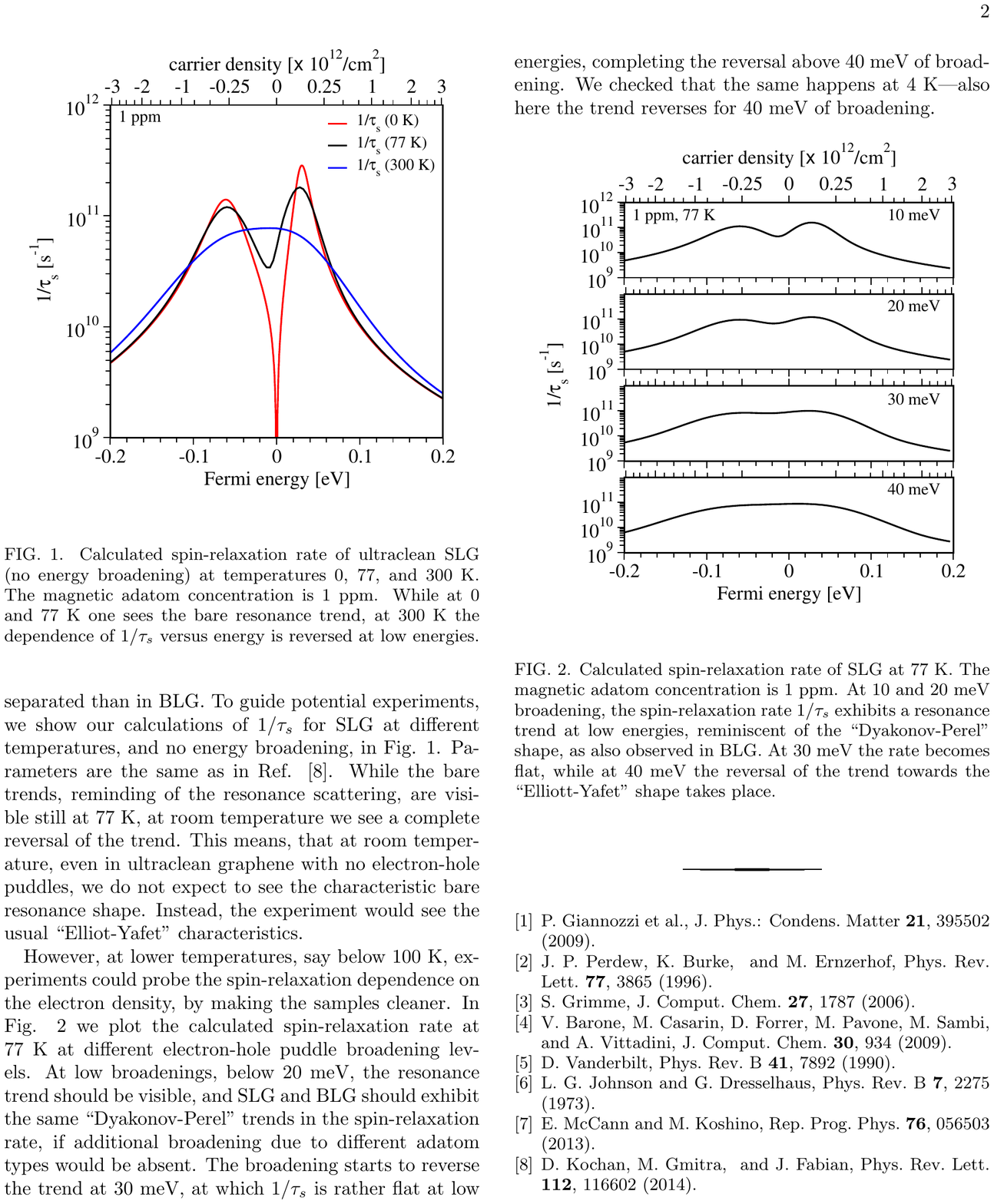}
\end{widetext}
%
%

\end{document}